\documentclass[prb,showpacs,preprint,nofloats,endfloats]{revtex4}

\usepackage{amsmath,amssymb,epsfig,epsf,psfig}

\begin{document}

\title
{\bf Pseudogaps: Introducing the Length Scale into DMFT}
\author{E.Z. Kuchinskii, I.A. Nekrasov, M.V. Sadovskii}

\affiliation
{Institute for Electrophysics, Russian Academy of Sciences,
Ekaterinburg, 620016, Russia}

\begin{abstract}
Pseudogap physics in strongly correlated systems is essentially scale
dependent. We generalize the dynamical--mean field theory (DMFT) 
by including into the DMFT equations dependence on correlation length of 
pseudogap fluctuations via additional (momentum dependent) self--energy 
$\Sigma_{\bf k}$.  This self -- energy describes non-local dynamical 
correlations induced by short--ranged collective SDW--like antiferromagnetic 
spin (or CDW--like charge) fluctuations.  At high enough temperatures these 
fluctuations can be viewed as a quenched Gaussian random field with finite 
correlation length.  This generalized DMFT+$\Sigma_{\bf k}$ approach is used 
for the numerical solution of the weakly doped one--band Hubbard model with 
repulsive Coulomb interaction on a square lattice with nearest and next 
nearest neighbour hopping.  The effective single impurity problem  
is solved by numerical renormalization group (NRG).  
Both types of strongly correlated metals, namely (i) doped Mott 
insulator and (ii) the case of bandwidth $W\lesssim U$ ($U$ --- value of 
local Coulomb interaction) are considered. Densities of states, spectral 
functions and ARPES spectra calculated within DMFT+$\Sigma_{\bf k}$ show a 
pseudogap formation near the Fermi level of the quasiparticle band.  We also
briefly discuss effects of random impurity scattering. Finally we demonstrate 
the qualitative picture of Fermi surface ``destruction'' due to pseudogap
fluctuations and formation of ``Fermi arcs'' which agrees well with ARPES 
observations.  
\end{abstract}


\maketitle

\newpage

\section{Introduction}

Pseudogap formation in the electronic spectrum of underdoped copper oxides 
\cite{Tim,MS} is especially striking anomaly of the normal state of
high temperature superconductors.  
Despite continuing discussions on the nature of
the pseudogap, we believe that the preferable ``scenario'' for its formation 
is most likely based on the model of strong scattering of the charge
carriers by short--ranged antiferromagnetic (AFM, SDW) spin fluctuations
\cite{MS,Pines}. In momentum representation this scattering transfers 
momenta of the order of ${\bf Q}=(\frac{\pi}{a},\frac{\pi}{a})$ 
($a$ --- lattice constant of two dimensional lattice). 
This leads to the formation of structures in the one-particle spectrum, 
which are precursors of the changes in the spectra due
to long--range AFM order (period doubling).
As a result we obtain non--Fermi liquid like behavior (dielectrization)
of the spectral density in the vicinity of the so called ``hot spots'' on the
Fermi surface, appearing at intersections of the Fermi surface 
with antiferromagnetic Brillouin zone boundary \cite{MS}.

Within this spin--fluctuation scenario a simplified model of the pseudogap 
state was studied \cite{MS,Sch,KS} under the assumption that the scattering
by dynamic spin fluctuations can be reduced for high enough temperatures
to a static Gaussian random field (quenched disorder) of pseudogap fluctuations.
These fluctuations are defined by a characteristic scattering vector from the 
vicinity of ${\bf Q}$,  with a width determined by the inverse correlation 
length of short--range order $\kappa=\xi^{-1}$. 

Undoped cuprates are antiferromagnetic Mott insulators with
$U\gg W$ ($U$ --- value of local Coulomb interaction, $W$ --- bandwidth of
non--interacting band), so that correlation effects are actually very important.  
It is thus clear that the electronic properties of 
underdoped (and probably also optimally doped) cuprates are governed by 
strong electronic correlations too, so that these systems are typical 
strongly correlated metals. Two types of correlated metals can be 
distinguished:  (i) the doped Mott insulator and (ii) the bandwidth 
controlled correlated metal $W\approx U$. 

A state of the art tool to describe such correlated  systems
is the dynamical mean--field theory (DMFT)
\cite{MetzVoll89,vollha93,pruschke,georges96,PT}.
The characteristic features of correlated systems within the DMFT
are the formation of incoherent structures, the so-called Hubbard bands,
split by the Coulomb interaction $U$, and a
quasiparticle (conduction) band near the Fermi level dynamically 
generated by the local correlations 
\cite{MetzVoll89,vollha93,pruschke,georges96,PT}.

Unfortunately, the DMFT is not useful to the study the
``antiferromagnetic'' scenario of pseudogap formation in strongly
correlated metals. This is due to the basic approximation of the DMFT, which
amounts to the complete neglect of non--local dynamical correlation effects.
The aim of the present paper is to describe the main results of a
semiphenomenological approach, formulated by us recently to overcome this 
difficulty \cite{cm05}.

The paper is organized as follows.
In section \ref{leng_intro} we present a formulation of the
self--consistent generalization we call DMFT+$\Sigma_{\bf k}$ 
which includes short--ranged (non--local) correlations.
Section \ref{kself} describes the construction of the
{\bf k}--dependent self--energy, and some 
computational details are presented in section \ref{compdet}.  
Results and a discussion are given in the sections \ref{results} and 
\ref{concl}.

\section{Introducing length scale into DMFT: DMFT+$\Sigma_{\bf k}$ approach}
\label{leng_intro}

Basic shortcoming of traditional DMFT approach
\cite{MetzVoll89,vollha93,pruschke,georges96,PT}
is the neglect of momentum dependence of electron self--energy.
This approximation in principle allows for an exact solution of correlated
electron systems (in infinite dimensions) fully preserving the local part of 
the dynamics introduced by electronic correlations.
To include non--local effects, while remaining within the usual ``impurity
analogy''of DMFT, we propose the following procedure. 
To be definite, let us consider
a standard one-band Hubbard model. The extension to multi--orbital or
multi--band models is straightforward.
The major assumption of our approach is that the lattice
and Matsubara ``time'' Fourier transform of the single-particle Green function 
can be written as:
\begin{equation}
G_{\bf k}(i\omega)=\frac{1}{i\omega+\mu-\varepsilon({\bf k})-\Sigma(i\omega)
-\Sigma_{\bf k}(i\omega)},\qquad \omega=\pi T(2n+1),
\label{Gk}
\end{equation}
where $\Sigma(i\omega)$ is the {\em local} contribution to self--energy,
surviving in the DMFT, while $\Sigma_{\bf k}(i\omega)$
is some momentum dependent part. We suppose that
this last contribution is due to either electron interactions with some
``additional'' collective modes or order parameter fluctuations, or may be
due to similar non--local contributions within the Hubbard model itself. 
To avoid possible confusion we must stress that $\Sigma_{\bf k}(i\omega)$
can also contain  local (momentum independent) contribution which obviously
{\em vanishes} in the limit of infinite dimensionality $d\to\infty$ and is
not taken into account within the standard DMFT. 
Due to this fact there is no double
counting of diagrams problem within our approach for the Hubbard model.
This question does not arise at all if we consider $\Sigma_{\bf k}(i\omega)$
appearing due to some ``additional'' interaction. More important is that the 
assumed additive form of self--energy $\Sigma(i\omega)+\Sigma_{\bf k}(i\omega)$
implicitly corresponds to neglect of possible interference
of these local (DMFT) and non--local contributions.

The self--consistency equations of our generalized DMFT+$\Sigma_{\bf k}$ 
approach are formulated as follows \cite{cm05}:
\begin{enumerate}
\item{Start with some initial guess of {\em local} self--energy
$\Sigma(i\omega)$, e.g. $\Sigma(i\omega)=0$}.  
\item{Construct $\Sigma_{\bf k}(i\omega)$ within some (approximate) scheme, 
taking into account interactions with collective modes or order parameter
fluctuations which in general can depend on $\Sigma(i\omega)$
and $\mu$.} 
\item{Calculate the local Green function  
\begin{equation}
G_{ii}(i\omega)=\frac{1}{N}\sum_{\bf k}\frac{1}{i\omega+\mu
-\varepsilon({\bf k})-\Sigma(i\omega)-\Sigma_{\bf k}(i\omega)}.
\label{Gloc}
\end{equation}
}
\item{Define the ``Weiss field''
\begin{equation}
{\cal G}^{-1}_0(i\omega)=\Sigma(i\omega)+G^{-1}_{ii}(i\omega).
\label{Wss}
\end{equation}
}
\item{Using some ``impurity solver'' to calculate the single-particle Green 
function for the effective Anderson impurity problem, defined by 
Grassmanian integral 
\begin{equation}
G_{d}(\tau-\tau')=\frac{1}{Z_{\text{eff}}}
\int Dc^+_{i\sigma}Dc_{i\sigma}
c_{i\sigma}(\tau)c^+_{i\sigma}(\tau')\exp(-S_{\text{eff}})
\label{AndImp}
\end{equation}
with effective action for a fixed site (``impurity'') $i$
\begin{equation}
S_{\text{eff}}=-\int_{0}^{\beta}d\tau_1\int_{0}^{\beta}
d\tau_2c_{i\sigma}(\tau_1){\cal G}^{-1}_0(\tau_1-\tau_2)c^+_{i\sigma}(\tau_2)
+\int_{0}^{\beta}d\tau Un_{i\uparrow}(\tau)n_{i\downarrow}(\tau)\;\;,
\label{Seff}
\end{equation}
$Z_{\text{eff}}=\int Dc^+_{i\sigma}Dc_{i\sigma}\exp(-S_{\text{eff}})$, and
$\beta=T^{-1}$. This step produces a {\em new} set of values 
$G^{-1}_{d}(i\omega)$.}
\item{Define a {\em new} local self--energy
\begin{equation}
\Sigma(i\omega)={\cal G}^{-1}_0(i\omega)-G^{-1}_{d}(i\omega).
\label{StS}
\end{equation}
}
\item{Using this self--energy as ``initial'' one in step 1, continue the 
procedure until (and if) convergency is reached to obtain
\begin{equation}
G_{ii}(i\omega)=G_{d}(i\omega).
\label{G00}
\end{equation}
}
\end{enumerate}
Eventually, we get the desired Green function in the form of (\ref{Gk}),
where $\Sigma(i\omega)$ and $\Sigma_{\bf k}(i\omega)$ are those appearing
at the end of our iteration procedure.

\section{Construction of {\bf k}--dependent self--energy}
\label{kself}
For the momentum dependent part of the single-particle self--energy we 
concentrate on the effects of scattering of electrons from collective 
short-range SDW--like antiferromagnetic spin (or CDW--like charge) 
fluctuations.  To calculate $\Sigma_{\bf k}(i\omega)$ for an electron moving 
in the quenched random field of (static) Gaussian spin (or charge) 
fluctuations with dominant scattering momentum transfers from the vicinity of 
some characteristic vector ${\bf Q}$  (``hot spots'' model \cite{MS}), we use 
a slightly generalized version of the recursion procedure proposed in 
Refs.~\cite{MS79,Sch,KS} which takes into account {\em all} Feynman diagrams 
describing the scattering of electrons by this random field.  This becomes 
possible due to a remarkable property of our simplified version of ``hot 
spots'' model that  {\em the contribution of an arbitrary diagram with 
intersecting interaction lines is actually equal to the contribution of some 
diagram of the same order without intersections of these lines} 
\cite{MS79,KS}. 
Thus, in fact we can limit ourselves to consideration of only diagrams 
without intersecting interaction lines, taking the contribution of diagrams 
with intersections into account with the help of additional combinatorial 
factors, which are attributed to ``initial'' vertices or just interaction 
lines \cite{MS79}. As a result we obtain the following recursion relation 
(continuous fraction representation \cite{MS79}) for the desired 
self--energy:  
\begin{equation} 
\Sigma_{\bf k}(i\omega)=\Sigma_{n=1}(i\omega{\bf k}) 
\label{Sk} 
\end{equation} 
with 
\begin{equation}
\Sigma_{n}(i\omega{\bf k})=\Delta^2\frac{s(n)}
{i\omega+\mu-\Sigma(i\omega)
-\varepsilon_n({\bf k})+inv_n\kappa-\Sigma_{n+1}(i\omega{\bf k})}\;\;. 
\label{rec}
\end{equation} 
The quantity $\Delta$ characterizes the energy scale and
$\kappa=\xi^{-1}$ is the inverse correlation length of short range
SDW (CDW) fluctuations, $\varepsilon_n({\bf k})=\varepsilon({\bf k+Q})$ and 
$v_n=|v_{\bf k+Q}^{x}|+|v_{\bf k+Q}^{y}|$ 
for odd $n$ while $\varepsilon_n({\bf k})=\varepsilon({\bf k})$ and $v_{n}=
|v_{\bf k}^x|+|v_{\bf k}^{y}|$ for even $n$. The velocity projections
$v_{\bf k}^{x}$ and $v_{\bf k}^{y}$ are determined by usual momentum derivatives
of the ``bare'' electronic energy dispersion $\varepsilon({\bf k})$. Finally,
$s(n)$ represents a combinatorial factor with
\begin{equation}
s(n)=n
\label{vcomm}
\end{equation}
for the case of commensurate charge (CDW type) fluctuations with
${\bf Q}=(\pi/a,\pi/a)$ \cite{MS79}. 
For incommensurate CDW fluctuations \cite{MS79} one finds
\begin{equation} 
s(n)=\left\{\begin{array}{cc}
\frac{n+1}{2} & \mbox{for odd $n$} \\
\frac{n}{2} & \mbox{for even $n$}.
\end{array} \right.
\label{vinc}
\end{equation}
If we take into account the (Heisenberg) spin structure of interaction with 
spin fluctuations in  ``nearly antiferromagnetic Fermi--liquid'' 
(spin--fermion (SF) model Ref.~\cite{Sch}),
the combinatorics of diagrams becomes more complicated.
Spin--conserving scattering processes obeys commensurate combinatorics,
while spin--flip scattering is described by diagrams of incommensurate
type (``charged'' random field in terms of Ref.~\cite{Sch}). In this model
the recursion relation for the single-particle Green function is again given 
by (\ref{rec}), but the combinatorial factor $s(n)$ now acquires the following 
form \cite{Sch}:
\begin{equation} 
s(n)=\left\{\begin{array}{cc}
\frac{n+2}{3} & \mbox{for odd $n$} \\
\frac{n}{3} & \mbox{for even $n$}.
\end{array} \right.
\label{vspin}
\end{equation}
Obviously, with this procedure we introduce an important length scale $\xi$ 
not present in standard DMFT. Physically this scale mimics the effect of 
short--range (SDW or CDW) correlations within fermionic ``bath'' surrounding 
the effective  Anderson impurity. We expect that such a length--scale 
dependence will lead to a competition between local and non-local physics.

An important aspect of the theory is that both parameters $\Delta$ and $\xi$ 
can in principle be calculated from the microscopic model at hand. 
For example, using the two--particle selfconsistent approach of 
Ref.~\cite{VT} with the approximations introduced in Refs.~\cite{Sch,KS}, one 
can derive \cite{cm05} within the standard Hubbard model the following 
microscopic expression for $\Delta$:  
\begin{eqnarray} 
\Delta^2=\frac{1}{4}U^2\frac{<n_{i\uparrow}n_{i\downarrow}>}
{<n_{i\uparrow}><n_{i\downarrow}>}[<n_{i\uparrow}>+<n_{i\downarrow}>
-2<n_{i\uparrow}n_{i\downarrow}>]=\nonumber\\
=U^2\frac{<n_{i\uparrow}n_{i\downarrow}>}{n^2}<(n_{i\uparrow}
-n_{i\downarrow})^2>
=\nonumber\\
=U^2\frac{<n_{i\uparrow}n_{i\downarrow}>}{n^2}\frac{1}{3}<{\vec S}_i^2>,
\label{DeltHubb}
\end{eqnarray}
where we consider only scattering from antiferromagnetic spin fluctuations.
Different local quantities here -- spin fluctuation $<{\vec S}_i^2>$,  density
$n$ and  double occupancy $<n_{i\uparrow}n_{i\downarrow}>$ -- 
can easily be calculated within the standard DMFT \cite{georges96}. 
We performed such calculations~\cite{cm05} 
for wide range of $U$ and filling factors $n$ using quantum Monte--Carlo 
(QMC)~\cite{QMC}. From these calculations we can see that the values of
$\Delta$ lie in the interval of $\Delta=(0.5 \div 2.0)t$ and change rather
smoothly with $n$ and $U$.

Microscopic expressions for the correlation length 
$\xi=\kappa^{-1}$ can also be derived within the two--particle self--consistent
approach \cite{VT}. However, we expect those results for $\xi$ to be less 
reliable, because this approach is valid only for relatively small (or 
medium) values of $U/t$ and for purely two--dimensional case (while real
systems are quasi--two--dimensional). 

Thus, in the following we will consider 
both $\Delta$ and especially $\xi$ as some phenomenological parameters to be 
determined from experiments. This makes our approach somehow similar
in the spirit to Landau approach to Fermi--liquids.

Our construction can be further generalized to include other types of
interactions. Thus scattering by random impurities with point -- like 
potential $V$ is easily taken into account in self -- consistent Born 
approximation \cite{KKS}. Then, in comparison with impurity free case, 
just we have a substitution (renormalization):
\begin{eqnarray}
\varepsilon_n\to\varepsilon_n-\rho V^2\sum_{\bf p}ImG(\varepsilon_n{\bf p})
\equiv\varepsilon_n\eta_{\epsilon}
\label{reneps}
\\
\eta_{\epsilon}=1-\frac{\rho V^2 }{\varepsilon_n} \sum_{\bf p}ImG(\varepsilon_n{\bf p})
\label{etaeps}
\end{eqnarray}
If we do not perform fully self -- consistent calculations of impurity
self -- energy, in the simplest approximation we just have:
\begin{eqnarray}
\varepsilon_n\to\varepsilon_n\eta_{\epsilon} 
=\varepsilon_n+\gamma sign\varepsilon_n
\label{renepsi} 
\\ 
\eta_{\epsilon}=1+\frac{\gamma}{|\varepsilon_n|}
\label{etaepsi}
\end{eqnarray}
where $\gamma=\pi\rho V^2N_0(0)$ is the standard Born impurity scattering 
rate ($N_0(0)$ is the density of states of ``free'' electrons at the Fermi 
level).

\section{Results and discussion}
\label{results}

\subsection{Computation details}
\label{compdet}
In the following, we discuss results for a standard one-band
Hubbard model on a square lattice. With nearest ($t$) and
next nearest ($t'$) neighbour hopping integrals the dispersion reads
\begin{equation}
\varepsilon({\bf k})=-2t(\cos k_xa+\cos k_ya)-4t'\cos k_xa\cos k_ya\;\;,
\label{spectr}
\end{equation}
where $a$ is the lattice constant.
The correlations are introduced by a repulsive local two-particle interaction 
$U$. We choose as energy scale the nearest neighbour hopping integral $t$
and as length scale the lattice constant $a$.

For a square lattice the ``bare'' bandwidth is $W=8t$.
To study a strongly correlated metallic state obtained as doped Mott insulator
we use $U=40t$ as value for the Coulomb interaction and a filling $n=0.8$ (hole 
doping).  The correlated metal in the case of $W\gtrsim U$ 
is considered for the case of $U=4t$ and filling factor $n=0.8$ 
(hole doping). For $\Delta$ we have choosen rather typical values between 
$\Delta=0.1t$ and $\Delta=2t$ (actually as approximate limiting values 
obtained from (\ref{DeltHubb}) via QMC calculations in Ref.\cite{cm05}) 
and for the correlation length we considered mainly $\xi=2a$ and $\xi=10a$ 
(being motivated mainly by experimental data for cuprates~\cite{MS,Sch}).

The DMFT maps the lattice problem onto an effective, 
self--consistent impurity defined by Eqs. (\ref{AndImp})-(\ref{Seff}).  
In our work we employed as ``impurity solvers'' two 
reliable numerically exact methods --- quantum Monte--Carlo (QMC)~\cite{QMC} 
and numerical renormalization group (NRG) \cite{NRG,BPH}.
Calculations were done both for $t'=0$ and 
$t'/t$=-0.4 (more or less typical for cuprates)
at two different temperatures $T=0.088t$ and $T=0.356t$ (for NRG 
computations). QMC computations of double occupancies as functions of 
filling were done at temperatures $T=0.1t$ and $T=0.4t$. 

Below we present results only for most typical dependences and parameters,
more details can be found in Ref. \cite{cm05}.

\subsection{Generalized DMFT+$\Sigma_{\bf k}$ approach: densities of states}

\begin{figure}[htb]
\includegraphics[clip=true,width=0.45\textwidth]{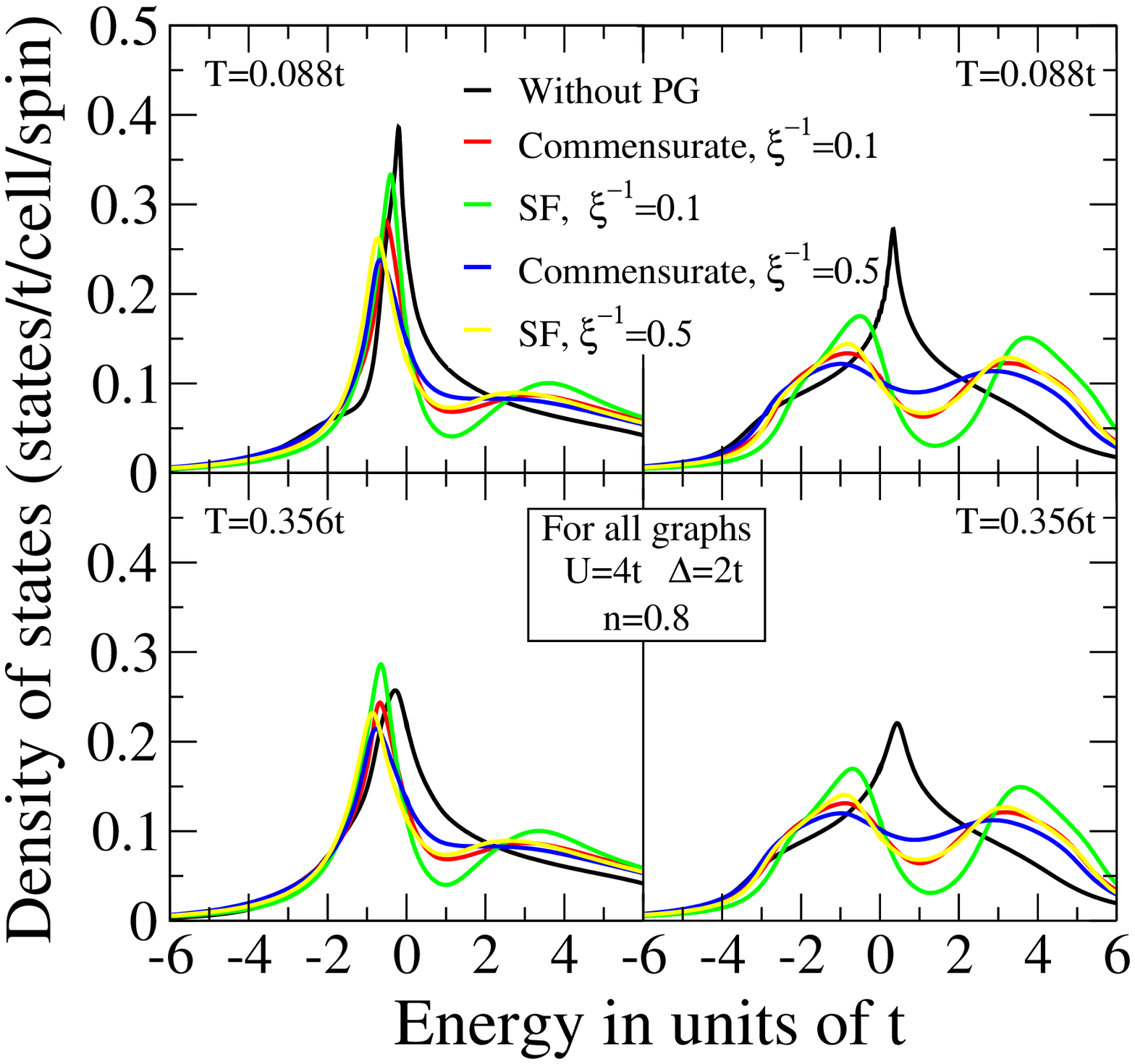}
\caption{Comparison of DOSs obtained from
DMFT(NRG)+$\Sigma_{\bf k}$ calculations for different
combinatorical factors (SF --- spin--fermion model, commensurate),
inverse correlation lengths
($\xi^{-1}$) in units of the lattice constant, temperatures ($T$) and 
pseudogap potential $\Delta=2t$.
Left column corresponds to $t'/t=-0.4$, right column to $t'=0$.
In all graphs the Coulomb interaction is $U=4t$ and $n=0.8$.
The Fermi level corresponds to zero.}
\label{DOS_4t_n08}
\end{figure}

Let us start the discussion of results
obtained within our generalized DMFT+$\Sigma_{\bf k}$ approach
with the densities of states (DOSs) for the case of small (relative
to bandwidth) Coulomb interaction $U=4t$ with and without pseudogap fluctuations.
As already discussed in the Introduction, the characteristic feature of the 
strongly correlated metallic state is the coexistence of lower and upper 
Hubbard bands split by the value of $U$ with a quasiparticle peak at the 
Fermi level.  Since at half--filling the bare DOS of the square lattice has a 
Van--Hove singularity at the Fermi level ($t'=0$) or close to it (in case of 
$t'/t=-0.4$) one cannot treat a peak on the Fermi level simply as a 
quasiparticle peak.  In fact, there are two contributions to this peak from 
(i) the quasiparticle peak appearing in strongly correlated metals due to 
many-body effects and (ii) the smoothed Van--Hove singularity from the bare 
DOS.  In Fig.~\ref{DOS_4t_n08} we show the corresponding DMFT(NRG) DOSs 
without pseudogap fluctuations as black lines for $n=0.8$ for both
bare dispersions $t'/t=-0.4$ (left panels) and for $t'=0$ (right panels) for two
different  temperatures $T=0.356t$ (middle panels) and $T=0.088t$ (upper and lower
panels).
The remaining curves in Fig.~\ref{DOS_4t_n08} represent results
for the DOSs with non-local fluctuations switched on.
For all sets of parameters one can see 
that the introduction of non-local fluctuations into the calculation leads to 
the formation of pseudogap within the quasiparticle peak.

For $n=0.8$ (Fig. \ref{DOS_4t_n08}) the picture of DOS is 
slightly asymmetric. The width of the pseudogap (the distance between peaks closest
to Fermi level) appears to be of the order of $\sim 2\Delta$. 
We have checked that decreasing the value of $\Delta$ from $2t$ to $t$ leads 
to a pseugogap that is correspondingly twice smaller and in addition more 
shallow. When one uses the combinatorial factors corresponding to the 
spin--fermion model (Eq.(\ref{vspin})), the pseudogap becomes more 
pronounced than in the case of commensurate charge fluctuations (combinatorial 
factors of Eq. (\ref{vinc})). 
The influence of the correlation length $\xi$ is also as expected. 
Changing $\xi^{-1}$ from $\xi^{-1}=0.1$ to $\xi^{-1}=0.5$, i.e.\ decreasing 
the range of the non-local fluctuations, slightly washes out the pseudogap.  
Also, increasing the temperature from $T=0.088t$ to $T=0.356t$ leads to a 
general broadening of the structures in the DOSs.  
Noteworthy is the fact that for $t'/t=-0.4$ and $\xi^{-1}=0.5$ the pseudogap 
has almost disappeared for the temperatures studied here.  
Also very remarkable point is the similarity of the results 
obtained with the generalized DMFT+$\Sigma_{\bf k}$ approach with $U=4t$ 
(smaller than the  bandwidth $W$) to those obtained earlier without 
Hubbard--like Coulomb interactions \cite{Sch,KS}.

Let us now consider the case of a doped Mott insulator.  The model parameters
are again taken as $t'/t=-0.4$ with filling factor of $n=0.8$, 
but the Coulomb interaction strength is set to $U=40t$.  
Characteristic features of the DOS for such a strongly correlated metal 
are a strong separation of lower and upper Hubbard bands and a Fermi level 
crossing by the lower Hubbard band (for non--half--filled case). 
Without non-local fluctuations the quasi--particle peak is again formed 
at the Fermi level, but now the upper Hubbard band is far to the 
right and does not touch the quasiparticle peak (as it was for the case of 
small Coulomb interactions). 

Pseudogap appears close to the middle of quasiparticle peak. In addition we 
observe that the lower Hubbard band is slightly broadened by fluctuation effects.  
Qualitative behaviour of the pseudogap anomalies is similar to those
described above for the case of $U=4t$, e.g.\ a decrease of $\xi$ makes the 
pseudogap less pronounced, while reducing $\Delta$ from $\Delta =2t$ to 
$\Delta =t$ narrows of the pseudogap and also makes it more shallow. 
Note that for the doped Mott--insulator  the pseudogap is 
remarkably more pronounced for the SDW--like fluctuations than for CDW--like 
fluctuations.

There are, however, quite clear differences to the case of $U=4t$. 
For example, the width of the pseudogap appears to be much smaller than 
$2\Delta$, which we attribute to the fact that the quasiparticle peak itself 
is actually rather narrow in the case of doped Mott insulator. 

\begin{figure}[htb]
\includegraphics[clip=true,width=0.4\textwidth,angle=270]{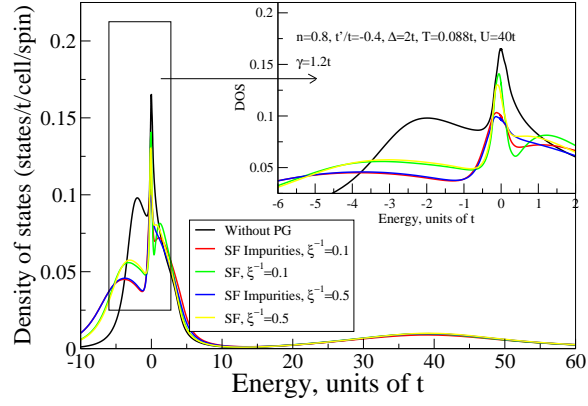}
\caption{Comparison of DOSs obtained from
DMFT(NRG)+$\Sigma_{\bf k}$ calculations for $t'/t=-0.4$, $T=0.088t$, $U=40t$, 
$n=0.8$ and for impurity scattering rate $\gamma=1.2t$.}
\label{dos_40t_04_imp}
\end{figure}

Random impurity scattering, in general case leads
to the filling of the pseudogap with the growth of impurity scattering rate
both for correlated metal and doped Mott insulator. As a typical example,
in Fig.~\ref{dos_40t_04_imp} we show results of our calculations for the case 
of doped Mott insulator. These were obtained via non self--consistent 
procedure (using (\ref{renepsi}), (\ref{etaepsi}), as full self--consistent 
procedure leads only to rather insignificant quantitative changes.

\subsection{Generalized DMFT+$\Sigma_{\bf k}$ approach:
spectral functions $A(\omega,{\bf k})$}

\begin{figure}[htb]
\includegraphics[clip=true,width=0.4\textwidth,angle=270]{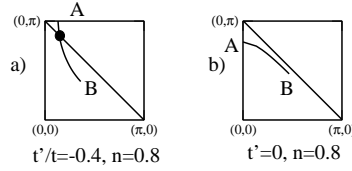}
\caption{1/8-th of the bare Fermi surfaces for
the different occupancies $n$ and combinations ($t,t'$)
used for the calculation of spectral functions $A({\bf k},\omega)$.
The diagonal line corresponds to the antiferromagnetic Brillouin zone
boundary at half--filling for a
square lattice with nearest-neighbours hopping only. The full circle marks 
the so-called ``hot--spot''.}
\label{FS_shapes}
\end{figure}

In the previous subsection we discussed the densities of states obtained
self--consistently by the DMFT+$\Sigma_{\bf k}$ approach. Once we get a
self--consistent solution of the DMFT+$\Sigma_{\bf k}$ equations with
non-local fluctuations we can, of course, also compute the spectral functions
$A(\omega,{\bf k})$
\begin{equation}
A(\omega,{\bf k})=-\frac{1}{\pi}{\rm Im}\frac{1}{\omega+\mu
-\varepsilon({\bf k})-\Sigma(\omega)-\Sigma_{\bf k}(\omega)},
\label{specf}
\end{equation}
where self--energy $\Sigma(\omega)$ and chemical potential $\mu$
are calculated self--consistently as described in Sec. \ref{leng_intro}. 
To plot $A(\omega,{\bf k})$ we choose ${\bf k}$--points along the
``bare'' Fermi surfaces for different types of
lattice spectra and fillings. In Fig. \ref{FS_shapes}
one can see corresponding shapes of these ``bare'' Fermi surfaces (presented 
are only 1/8-th parts of the Fermi surfaces within the first quadrant of the
Brillouin zone).

\begin{figure}[htb]
\includegraphics[clip=true,width=0.4\textwidth]{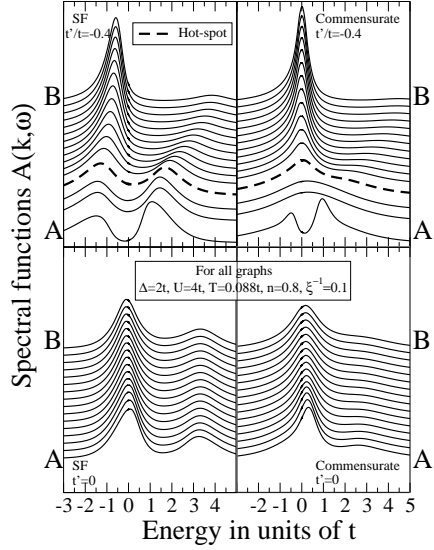}
\caption{Spectral functions $A({\bf k},\omega)$ 
obtained from the DMFT(NRG)+$\Sigma_{\bf k}$ calculations along the
directions shown in Fig. \ref{FS_shapes}. 
Model parameters were chosen as 
$U=4t$, $n=0.8$, $\Delta=2t$, $\xi^{-1}=0.1$ 
and temperature $T=0.088t$. 
The ``hot--spot'' {\bf k}-point is marked as fat dashed line.
The Fermi level corresponds to zero.}
\label{sf_U4t_n08}
\end{figure}

In the following we concentrate mainly on the case $U=4t$ and
filling $n=0.8$ (Fermi surface of Fig. \ref{FS_shapes}(a)). The
corresponding spectral functions $A(\omega,{\bf k})$ are depicted in
Fig.~\ref{sf_U4t_n08}. 
When $t'/t=-0.4$ (upper row), the spectral function close to the Brillouin zone 
diagonal (point B) has the typical Fermi--liquid behaviour, consisting of a 
rather sharp peak close to the Fermi level.  In the case of SDW--like 
fluctuations this peak is shifted down in energy by about $-0.5t$ (left upper 
corner).  In the vicinity of the ``hot--spot'' the shape of $A(\omega,{\bf k})$ 
is completely modified. Now $A(\omega,{\bf k})$ becomes double--peaked 
and non--Fermi--liquid--like.  Directly at the ``hot--spot'', 
$A(\omega,{\bf k})$ for SDW--like  fluctuations has two equally intensive 
peaks situated symmetrically around the Fermi level and split from each other 
by $\sim 1.5\Delta$ Refs.~\cite{Sch,KS}.  
For commensurate CDW--like fluctuations the
spectral function in the ``hot--spot'' region has one broad peak centered at 
the Fermi level with width $\sim \Delta$. 
Such a merging of the two peaks at the ``hot--spot'' 
for commensurate fluctuations was previously observed in Ref. \cite{KS}.
However close to point A  this type of fluctuations also produces a
double--peak structure in the spectral function.

In the lower panel of Fig.~\ref{sf_U4t_n08} we show spectral functions
hole doping ($n=0.8$) and the case of $t'=0$ (Fermi surface from 
Fig. \ref{FS_shapes}(b)). Since the Fermi surface now is everywhere close to 
the antiferromagnetic zone boundary, the pseudogap anomalies are rather 
strong and almost non--dispersive along the Fermi surface.  

\begin{figure}[htb]
\includegraphics[clip=true,width=0.4\textwidth]{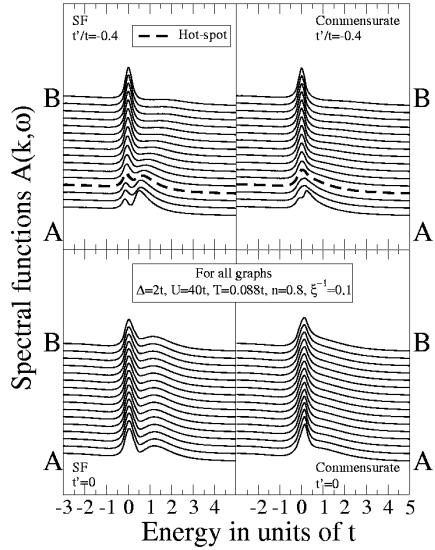}
\caption{Spectral functions $A({\bf k},\omega)$ 
obtained from the DMFT(NRG)+$\Sigma_{\bf k}$ calculations
for $U=40t$,
other parameters as in Fig.~\ref{sf_U4t_n08}.}
\label{sf_U40t_n08}
\end{figure}

For the case of a doped Mott insulator ($U=40t$, $n=0.8$), 
the spectral functions obtained by the DMFT+$\Sigma_{\bf k}$ approach are 
presented in Fig.~\ref{sf_U40t_n08}.  Qualitatively, the shapes of these 
spectral functions are similar to those shown on Fig.~\ref{sf_U4t_n08}.  As 
was pointed out above, the strong Coulomb correlations lead to a narrowing of 
the quasiparticle peak and a corresponding decrease of the pseudogap width. 
One should also note that in contrast to $U=4t$ the spectral functions are 
now less intensive, because part of the spectral weight is transferred to the 
upper Hubbard band.

\begin{figure}[htb]
\includegraphics[clip=true,width=0.4\textwidth]{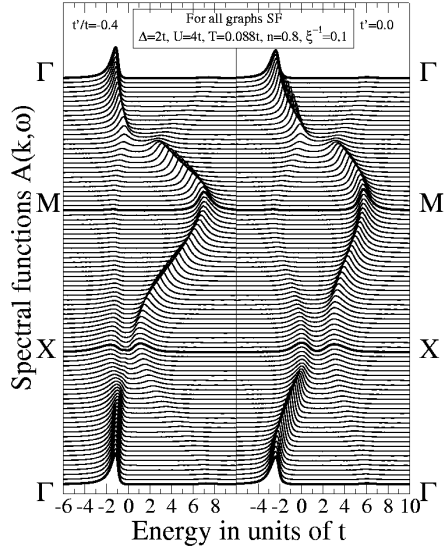}
\caption{Spectral functions $A({\bf k},\omega)$ obtained from the
DMFT(NRG)+$\Sigma_{\bf k}$ calculations
along high-symmetry directions of first Brilloin zone
$\Gamma(0,0)\!-\!\rm{X}(\pi,0)\!-\!\rm{M}(\pi,\pi)\!-\!\Gamma(0,0)$,
SF combinatorics (left row) and commensurate combinatorics (right column).
Other parameters are $U=4t$, $n=0.8$, $\Delta=2t$, $\xi^{-1}=0.1$ 
and temperature $T=0.088t$.
The Fermi level corresponds to zero.}
\label{n08_U4t_tri}
\end{figure}

Using another quite common choice of ${\bf k}$--points we can compute 
$A(\omega,{\bf k})$ along high--symmetry directions in the first 
Brillouin zone: 
$\Gamma(0,0)\!-\!\rm{X}(\pi,0)\!-\!\rm{M}(\pi,\pi)\!-\!\Gamma(0,0)$.  
The spectral functions for these 
${\bf k}$--points are shown in Fig.~ \ref{n08_U4t_tri} 
for the case of SDW--like fluctuations and $U=4t$.
For all sets of parameters one can see a characteristic double -- peak 
pseudogap structure close to the $X$ point. In the middle of 
$M\!-\!\Gamma$ direction (so called ``nodal'' point) one can see the 
reminiscence of AFM gap which has its biggest value here in case of perfect 
antiferromagnetic ordering. Also in the nodal point ``kink''-like
behavior is observed caused by interactions between correlated electrons with 
pseudo-gap fluctuations. A change of the filling leads mainly to a 
rigid shift of spectral functions with respect to the Fermi level. 
For the case of $U=40t$ spectral densities demonstrate rather similar
behavior \cite{cm05}.

\begin{figure}
\includegraphics[clip=true,width=0.4\textwidth]{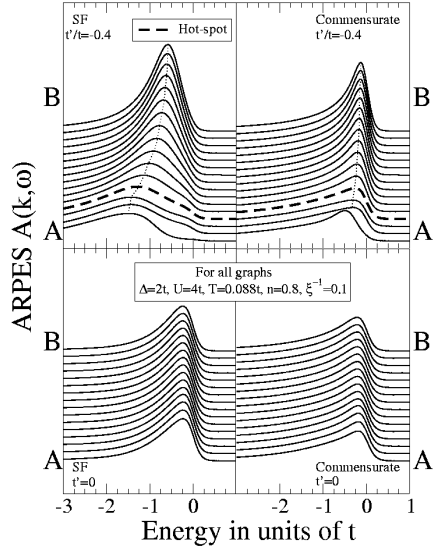}
\caption{ARPES  spectra obtained from the
DMFT(NRG)+$\Sigma_{\bf k}$ calculations for $U=4t$ and $n=0.8$
along the lines in the first BZ as depicted by
Fig.~\ref{FS_shapes}, all other parameters as in Fig.~\ref{sf_U4t_n08}.}
\label{arpes_U4t_n08}
\end{figure}

With the spectral functions we are now in a position to calculate
angle resolved photoemission spectra (ARPES), which is the most direct
experimental way to observe pseudogap in real compounds. 
For that purpose, we only need to multiply our results for the spectral 
functions with the Fermi function at appropriate temperature. 
Typical example of the resulting DMFT+$\Sigma_{\bf k}$ ARPES spectra 
are presented in Fig.~\ref{arpes_U4t_n08}.  
One should note that for $t'/t=-0.4$ (upper panel 
of Fig.~\ref{arpes_U4t_n08}) as ${\bf k}$ goes from point ``A'' to point 
``B'' the peak situated slightly below the Fermi level changes its position 
and moves down in energy.  Simultaneously it becomes more broad and less 
intensive.  The dotted line guides the motion of the peak maximum.  
Such behaviour of the peak in the ARPES is 
rather reminiscent of those observed experimentally in underdoped cuprates 
\cite{MS,Sch,Kam}.

\subsection{``Destruction'' of the Fermi surface}

Within the standard DMFT approach Fermi surface is not renormalized by
interactions and just coincides with that of the ``bare'' quasiparticles
\cite{vollha93}. However, in the case of nontrivial momentum dependence
of electron self -- energy, important renormalization of the Fermi surface
appears due to pseudogap formation \cite{Sch}.
There are a number of ways to define Fermi surface in strongly correlated
system with pseudogap fluctuations. In the following we are using intensity
plots (within the Brillouin zone) of the spectral density (\ref{specf}) 
taken at $\omega=0$. These are readily measured by ARPES and appropriate 
peak positions define the Fermi surface in the usual Fermi liquid case. 

\begin{figure}
\includegraphics[clip=true,width=0.65\textwidth]{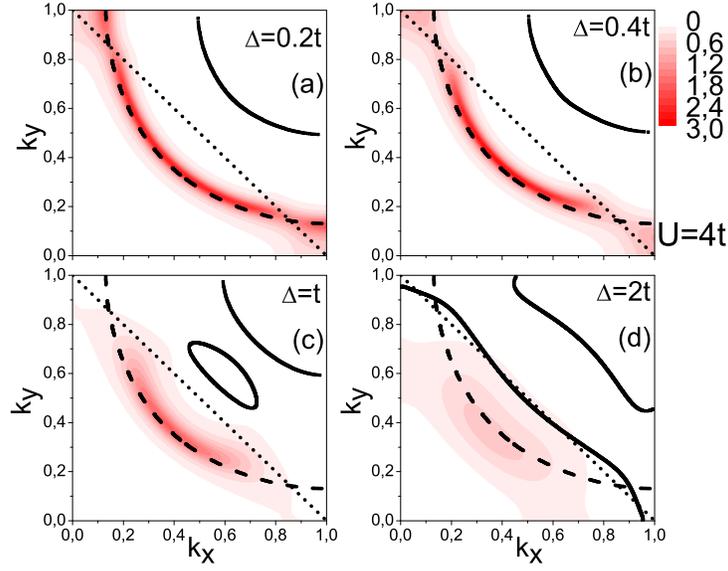}
\caption{Fermi surfaces obtained from the
DMFT(NRG)+$\Sigma_{\bf k}$ calculations for $U=4t$ and $n=0.8$.
Shown are intensity plots of spectral density (\ref{specf}) for 
$\omega=0$ and $\xi=10a$.
(a) -- $\Delta=0.2t$;\ (b) -- $\Delta=0.4t$;\ (c) -- $\Delta=t$;\
(d) -- $\Delta=2t$.
Dashed line denotes ``bare'' Fermi surface.
Black lines show the solution of Eq. (\ref{ReFS}).}
\label{FS_U4}
\end{figure}

\begin{figure}
\includegraphics[clip=true,width=0.65\textwidth]{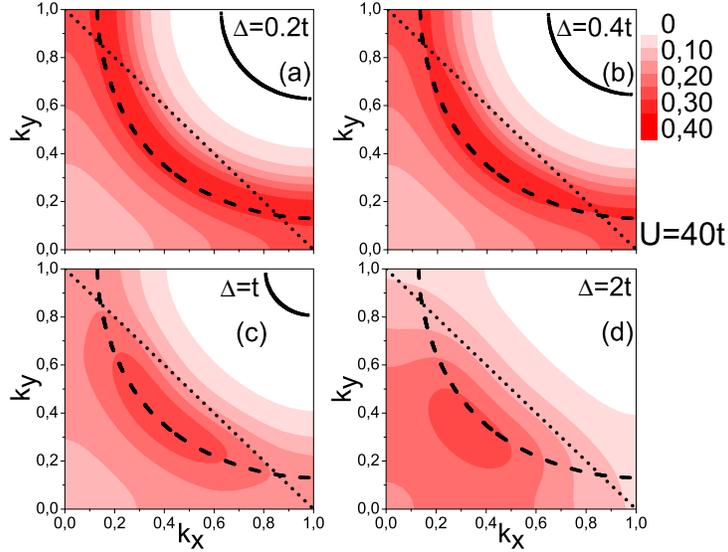}
\caption{``Destruction''of the Fermi surface obtained from the
DMFT(NRG)+$\Sigma_{\bf k}$ calculations for $U=40t$ and $n=0.8$.
Other parameters and notations are the same as in Fig.~\ref{FS_U4}.}
\label{FS_U40}
\end{figure}

Our results \cite{KNS} are shown in Fig.~\ref{FS_U4} for the case of correlated 
metal with $U=4t$ and in Fig.~\ref{FS_U40} for the doped Mott insulator 
($U=40t$) (in both cases we assume spin -- fermion combinatorics). 
The qualitative behavior observed in Fig.~\ref{FS_U4} clearly demonstrates
the ``destruction'' of the well defined Fermi surface in the strongly
correlated metal with the growth of the pseudogap amplitude $\Delta$. 
Quite similar behavior was first observed
in pioneering paper by Norman et al.\cite{Norm} and in numerous later ARPES 
experiments. It is seen, that ``destruction'' of the Fermi surface starts
in the vicinity of ``hot spots'' for small values of $\Delta$, but almost
immediately it disappears in the whole antinodal region of the Brillouin
zone, while only ``Fermi arcs'' remain in the nodal region very close to the
``bare'' Fermi surface. These results
give a natural explanation of the observed behavior and also of the fact that
the existence of ``hot spots'' regions was observed only in some rare cases
\cite{Arm}. 

For the case of doped Mott insulator shown in Fig.~\ref{FS_U40}
we see that the Fermi surface is rather poorly defined for all values of
$\Delta$, as the spectral density profiles are much more
``blurred'' than in the case of smaller values of $U$, reflecting important
role of correlations. 

It is interesting to note that from Figs.~\ref{FS_U4},
\ref{FS_U40} it is clearly seen that rather natural definition
of the Fermi surface as defined by the solution of the equation
\begin{equation}
\omega-\varepsilon({\bf k})+\mu-Re\Sigma(\omega)-Re\Sigma_{\bf k}(\omega)=0
\label{ReFS}
\end{equation}
for $\omega=0$, used e.g. in Ref.\cite{Sch} is inadequate for strongly
correlated systems with finite $U$ and nonlocal interactions (pseudogap
fluctuations). 

\section{Conclusion}
\label{concl}

To summarize, we propose a generalized DMFT+$\Sigma_{\bf k}$
approach, which is meant to take into account the important
effects due to non--local correlations in a systematic, but to some extent
phenomenological fashion.
The main idea of this extension is to stay within a
usual effective Anderson impurity analogy, and introduce length scale
dependence due to non-local correlation via the effective medium (``bath'') 
appearing in the standard DMFT. 
This becomes possible by incorporating scattering processes of fermions
in the ``bath'' from non--local collective SDW--like antiferromagnetic spin 
(or CDW--like charge) fluctuations.  Such a generalization of the DMFT allows 
one to overcome the well--known shortcoming of ${\bf k}$--independence of 
self--energy of the standard DMFT.  It in turn opens the possibility to 
access the physics of low--dimensional strongly correlated systems, where 
different types of spatial fluctuations (e.g. of some order parameter), 
become important.  However, we must stress that our 
procedure in no way introduces any kind of systematic $1/d$--expansion, being 
only a qualitative method to include length scale into DMFT.

In our present study we addressed the problem of pseudogap formation in the
strongly correlated metallic state.  We showed evidence that the pseudogap 
appears at the Fermi level within the quasiparticle peak, introducing a new
small energy scale of the order of $\Delta$ in the DOSs and spectral functions
$A(\omega,{\bf k})$ and significant renormalization of the Fermi surface.

Let us stress, that our generalization of DMFT leads to non--trivial 
and in our opinion physically sensible ${\bf k}$--dependence of spectral 
functions. Similar results were obtained in recent years using the cluster mean-field 
theories \cite{TMrmp}. The major advantage of our approach over these 
theories is, that we stay in an effective single-impurity 
picture. This means that our approach is computationally much less expensive 
and therefore also easily generalizable for the account of additional
interactions .

\section{Acknowledgements}

We are grateful to Th. Pruschke for providing us with his NRG code and
helpful discussions.
This work was supported in part by RFBR grants 05-02-16301, 05-02-17244, 
and programs of the
Presidium of the Russian Academy of Sciences (RAS) ``Quantum macrophysics''
and of the Division of Physical Sciences of the RAS ``Strongly correlated
electrons in semiconductors, metals, superconductors and magnetic
materials''. I.N. acknowledges support
from the Dynasty Foundation and International
Centre for Fundamental Physics in Moscow program for young
scientists 2005 and Russian Science Support Foundation program for
young PhD of the Russian Academy of Sciences 2005.

\pagestyle{empty}

\newpage

\begin{figure}[htb]
\includegraphics[clip=true,width=0.4\textwidth,angle=270]{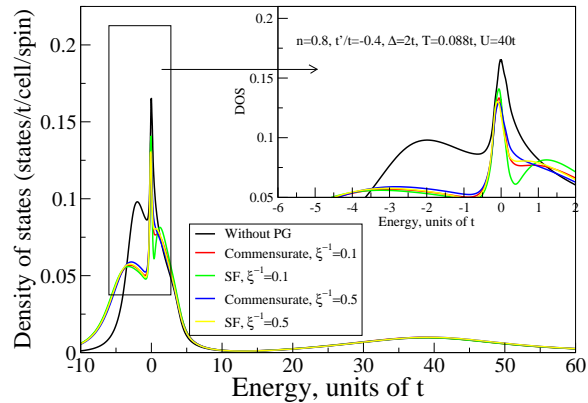}
\caption{Comparison of DOSs obtained from
DMFT(NRG)+$\Sigma_{\bf k}$ calculations for $t'/t=-0.4$, $T=0.088t$, $U=40t$ and filling $n=0.8$.}
\label{dos_40t_04}
\end{figure}

\end{document}